\documentclass[%
reprint,
superscriptaddress,
 amsmath,amssymb,
 aps,
]{revtex4-2}

\usepackage{graphicx}
\usepackage{bm, physics}
\usepackage{float}
\usepackage{hyperref}

\begin{document}

\title{Dynamics of measurement-induced state transitions in superconducting qubits}
\author{Yuta Hirasaki}
 \affiliation{ 
Department of Applied Physics, The University of Tokyo, Tokyo 113-8656, Japan.
}%
\author{Shunsuke Daimon}%
 \email{daimon.shunsuke@qst.go.jp}
 \affiliation{ 
Department of Applied Physics, The University of Tokyo, Tokyo 113-8656, Japan.
}%
 \affiliation{ 
Quantum Materials and Applications Research Center, National Institutes for Quantum Science and Technology, Tokyo 152-8550, Japan.
}%
\author{Naoki Kanazawa}
\author{Toshinari Itoko}
\author{Masao Tokunari}
\affiliation{%
IBM Quantum, IBM Research Tokyo, 19-21 Nihonbashi Hakozaki-cho, Chuo-ku, Tokyo, 103-8510, Japan.
}%
\author{Eiji Saitoh}
\affiliation{ 
Department of Applied Physics, The University of Tokyo, Tokyo 113-8656, Japan.
}%
\affiliation{Institute for AI and Beyond, The University of Tokyo, Tokyo 113-8656, Japan.}
\affiliation{WPI Advanced Institute for Materials Research, Tohoku University, Sendai 980-8577, Japan.}
\affiliation{RIKEN Center for Emergent Matter Science (CEMS), Wako 351-0198, Japan.}

\date{\today}

\begin{abstract}
We have investigated temporal fluctuation of superconducting qubits via the time-resolved measurement for an IBM Quantum system. We found that the qubit error rate abruptly changes during specific time intervals. Each high error state persists for several tens of seconds, and exhibits an on-off behavior. The observed temporal instability can be attributed to qubit transitions induced by a measurement stimulus. Resonant transition between fluctuating dressed states of the qubits coupled with high-frequency resonators can be responsible for the error-rate change.
\end{abstract}


\maketitle 
Quantum computers are expected to perform computational abilities that far exceed those of their classical counterparts~\cite{nielsen2010quantum}. Many researchers have been making tremendous efforts to develop technologies that apply the principles of quantum mechanics to information processing~\cite{krantz2019quantum, bruzewicz2019trapped, o2007optical, gaita2019molecular}. Diverse materials or artificial atoms can serve as qubits, such as trapped ions~\cite{HAFFNER2008155}, quantum dots~\cite{PhysRevA.69.042302}, and many others. Among these candidates, superconducting qubits~\cite{Wendin_2017} have gained significant attention for their relatively long coherence times and the ability to implement fast gate operations~\cite{kjaergaard2020superconducting, krantz2019quantum, 10.1063/5.0029735}. A superconducting qubit-resonator system is well described by the Jaynes-Cummings model~\cite{1443594}. Within this framework, quantum nondemolition measurements are achieved by using the dispersive interaction between the qubit and resonator photons. 

Existing superconducting quantum computers cannot escape from noise~\cite{siddiqi2021engineering, burnett2019decoherence}, including qubit decoherence~\cite{PhysRevB.86.184514} and imperfect qubit operations~\cite{PhysRevLett.127.080505} and readout~\cite{PhysRevApplied.10.034040}. These factors ultimately limit the computational abilities and degrade the fidelity of qubit outputs. Several surveys have revealed that qubit errors show fluctuating behavior over time~\cite{klimov2018fluctuations, carroll2022dynamics, de2020two, hirasaki2023detection, PRXQuantum.4.020356}, which presents a significant challenge for current error mitigation techniques~\cite{temme2017error, kim2023scalable, van2023probabilistic}. It is imperative to investigate the error dynamics of superconducting qubit systems.

In previous reports~\cite{PhysRevLett.117.190503, PhysRevApplied.20.054008}, the mechanism of measurement-induced state transitions (MIST) has been extensively studied. It was found that when the measurement stimulus is injected into the readout cavity, qubit dressed states come into resonance and the qubit is excited beyond the computational subspace. In addition, it was reported that the resonance shows a noisy behavior when measured repeatedly, which can be attributed to a fluctuating offset charge. However, the early studies are limited to a system of superconducting qubits coupled with low-frequency readout resonators. In addition, the dynamics of MIST has not been fully investigated.

In this study, we conduct a time-resolved MIST experiment for qubits in an IBM Quantum system coupled with high-frequency readout resonators. We perform the experiment for a total duration of approximately $2.70\times 10^5$ seconds, and we analyze the occurrence rate and the duration of MIST. We probe how dressed state energy fluctuates over time and temporarily resonates with other energy levels.

All the experiments were performed on \texttt{ibm\_kawasaki}, an IBM Quantum system. This processor had 27 transmon qubits (this device has been updated and now it has 127 qubits.), and we used the qubit 7 in this processor. The qubit frequency is ${\omega_{01}}/{2\pi} = 5.457$ GHz, with the anharmonicity $\eta/2\pi = -336.6$ MHz, and the energy relaxation time is $104.9\;\mathrm{\mu s}$. The qubit is capacitively coupled to a readout resonator with its fundamental frequency $\omega_\mathrm{r}/2\pi = 7.117\;\mathrm{GHz}$, and the decay rate $\kappa = 1/(209\;\mathrm{ns})$. The coupling strength between the qubit and the resonator is $g/2\pi = 86.85\;\mathrm{MHz}$. The readout assignment error for this qubit is $1.46 \%$.

In this dispersive regime, we can measure the qubit quantum state by the dispersive readout. The dressed frequency of the resonator $\omega_{\mathrm{r}, \ket{i}}$ depends on the qubit quantum state $\ket{i}$, and we can infer the qubit state by probing the readout resonator with a measurement stimulus. The qubit information is encoded in the amplitude $A_\mathrm{out}$ and phase $\phi$ of the output microwave, as they are dependent on the resonator frequency. This signal is digitized and represented as a point in the in phase and quadrature (IQ) plane. 

\begin{figure}
\includegraphics{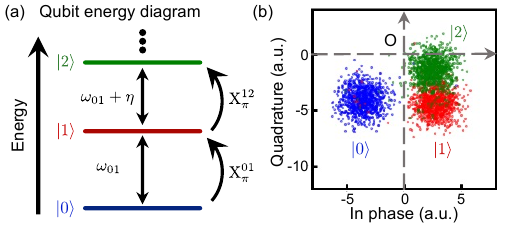}
\caption{
(a) A qubit energy diagram. The qubit is excited with the pulses whose frequencies are equal to the qubit transition frequencies. $\omega_{01}$ denotes the transition frequency between the $\ket{0}$ and the $\ket{1}$ states, $\eta$ denotes the anharmonicity, and $\mathrm{X}_{\pi}^{ij}$ denotes the $\pi$ pulse between the $\ket{i}$ and $\ket{j}$ states. (b) The in phase and quadrature (IQ) components of the output microwaves with the three different qubit states. The qubit is prepared in either the ground ($\ket{0}$) state, the first excited state ($\ket{1}$) or the second excited ($\ket{2}$) state. The blue (red) dots show the obtained data corresponding to the qubit $\ket{0}$ ($\ket{1}$) state. The green dots correspond to the $\ket{2}$ state.
}
\label{f1}
\end{figure}

We initialized the qubit into three different states, the ground ($\ket{0}$), the first excited ($\ket{1}$), and the second excited ($\ket{2}$) state as shown in Fig. \ref{f1}(a). We then performed measurements to obtain three distinct clusters corresponding to the three quantum states. As shown in Fig. \ref{f1}(b), the blue dots correspond to the qubit $\ket{0}$ state, the red dots correspond to the qubit $\ket{1}$ state, and the green dots correspond to the qubit $\ket{2}$ state. We note that the dispersive shift and the quality factor of this resonator allow us to resolve up to the first three states. Signals corresponding to the states higher than the $\ket{2}$ state are expected to appear near those of the $\ket{2}$ state~\cite{PhysRevResearch.4.033027}. Hereafter we denote states higher than the $\ket{1}$ state as $\ket{\mathrm{HS}}$. 

In the following experiments, the IQ measurement was repeated at regular intervals of several hundred microseconds over a period of several tens of minutes. At each interval, the qubit is initialized into the ground state with a reset operation. We experimentally confirmed that the reset operation leaves the HS states unchanged and initializes only the first excited state. Since the IQ signals distribute as shown in Fig. \ref{f1}(b), we can discriminate the $\ket{0}$ and the other states by focusing on the sign of the in phase component. A negative (positive) in phase component corresponds to the signal being converted to 0 (1). Henceforth, we denote the probability of obtaining states other than the $\ket{0}$ state as $P_1$. We repeat the IQ measurements $L$ times, convert the signals to the binary data based on the in phase components, and then take the ensemble average of $N$ data to obtain a time series of $P_1$.

\begin{figure}
  \includegraphics{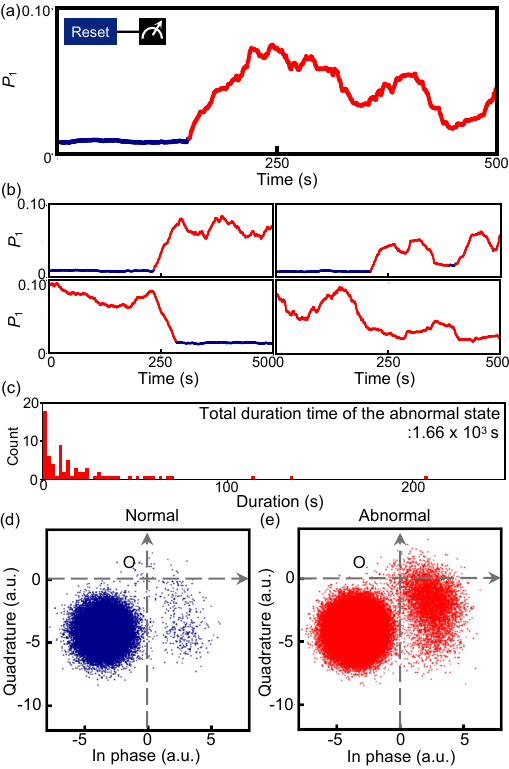}
  \caption{
  (a) Observed $P_1$ as a function of time. $P_1$ denotes the probability of measuring states other than the ground state. The inset to the top panel illustrates the used quantum circuit, where the qubit is initialized into the $\ket{0}$ state and then measured. We iterated the initialization and the IQ measurement every $500\;\mu\mathrm{s}$ for the entire $500$-second duration. 
  (b) The same experiment was conducted multiple times, and the resulting time series are illustrated in the four panels.
  (c) The histogram of the duration time of the observed anomaly. 
  (d)(e) A comparison of the IQ plots for the normal (blue) and abnormal (red) time periods indicated in (a). The left (right) panel corresponds to the normal (abnormal) time period.
  }
  \label{f2}
\end{figure}

We conducted the IQ measurement for 500 s and calculated the time series of $P_1$ with $N = 16384$ as shown in Fig. \ref{f2}(a). We interleaved several different quantum circuits to simulate practical quantum computation (see Appendix~\ref{tsp1} for the details of the experimental setup). In Fig. \ref{f2}(a), $P_1$ remains around $0.01$ until 230 s (indicated by the blue line), consistent with the measurement of the $\ket{0}$ state as shown in the inset. From 230 s to 470 s, however, $P_1$ shows a sudden rise to around $0.08$ (indicated by the red line). This sudden rise cannot be explained in terms of a sampling error since the standard deviation is given by $\sigma = \sqrt{\frac{P_1(1-P_1)}{N}}\approx 7.8\times 10^{-4}$. This behavior is frequently observed in multiple instances, as demonstrated in the four panels of Fig. \ref{f2}(b). We repeated this experiment for approximately $2.7\times 10^5$ seconds to inspect the endurance of the observed anomaly. The histogram of the duration time of the anomaly is shown in Fig. \ref{f2}(c). The duration varies from several seconds to around 200 seconds and the anomaly is repeatedly observed. The anomaly appeared for a total of $1.66 \times 10^3$ seconds throughout the entire experiments, which spanned $2.70\times 10^5$ seconds.

As a temporal increase in $P_1$ often results from a temporal signal rotation in the IQ plane, we compare the IQ plots of the two time periods in Fig. \ref{f2}(a). The result is shown in Fig. \ref{f2}(d) and (e). The left (right) panel shows the IQ signals in the lower (higher) error time period, as indicated by the blue (red) line of Fig. \ref{f2}(a). In the left panel, most of the dots are clustered around a center in the fourth quadrant, which indicates that the qubit is in the $\ket{0}$ state. However, in the right panel, we observe distinct dense signals near the origin in the first quadrant, suggesting that the qubit is excited to $\ket{\mathrm{HS}}$ states. Excitations beyond the qubit subspace contribute to the higher error time period depicted by the red line in Fig. \ref{f2}(a). 

\begin{figure}
  \includegraphics{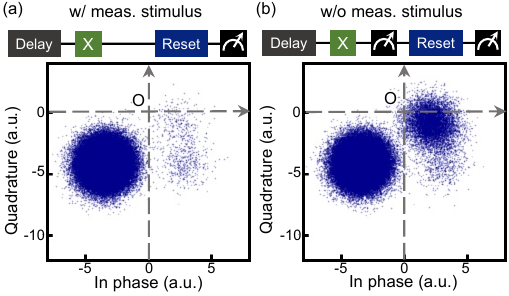}
  \caption{
  Controlled experiment to specify the excitation origin. (a) The qubit thermalizes to the ground state after the $1000\;\mathrm{\mu s}$ delay, and then it is excited to the $\ket{1}$ state with the $\pi$ pulse. The reset operation initialize the qubit $\ket{1}$ state into the $\ket{0}$ state, followed by the qubit measurement. The resulting signals are plotted in the IQ plane at the bottom. (b) The measurement stimulus is transmitted to the qubit after the qubit is excited to the $\ket{1}$ state. The obtained signals are shown at the bottom.
  }
  \label{f3}
\end{figure}

From the result in Fig. \ref{f2}, it can be deduced that the temporal HS excitation is triggered by either the reset operation or the measurement stimulus. To specify the underlying mechanism of the observed excitation, we conducted the experiment shown in Fig. \ref{f3}. In this controlled experiment, the difference lies in the measurement stimulus, allowing us to establish that any observed difference between the two experiments is triggered by the measurement stimulus. These two experiments are performed alternately with an interval of several hundred microseconds. In Fig. \ref{f3}(a), we first wait for $1000\;\mathrm{\mu s}$ to initializes the qubit into the thermal ground state, and apply a $\pi$ pulse to excite it to the $\ket{1}$ state. The $\ket{1}$ state is initialized into the $\ket{0}$ state with a reset operation, and the qubit is finally measured. The corresponding signals are plotted in the IQ plane at the bottom. In Fig. \ref{f3}(b), we insert transmitting the measurement stimulus after the qubit is excited to the $\ket{1}$ state with the $\pi$ pulse. The IQ signals obtained after the reset are plotted in the IQ plane at the bottom of Fig. \ref{f3}(b).

In Fig. \ref{f3}(b), a sharp density near the origin is observed, while the majority of the IQ signals are distributed in the fourth quadrant. This result shows that the excitation in Fig. \ref{f2} is triggered by transmitting the measurement stimulus to the qubit in the $\ket{1}$ state. Remarkably, transmitting the measurement stimulus to the qubit in the $\ket{0}$ state does not result in excitation to $\ket{\mathrm{HS}}$ states. These findings suggest that measuring the $\ket{1}$ state excites the qubit beyond the computational subspace. Note that this MIST is observed only during specific time periods, as discussed in Fig. \ref{f2}.

\begin{figure}
\includegraphics{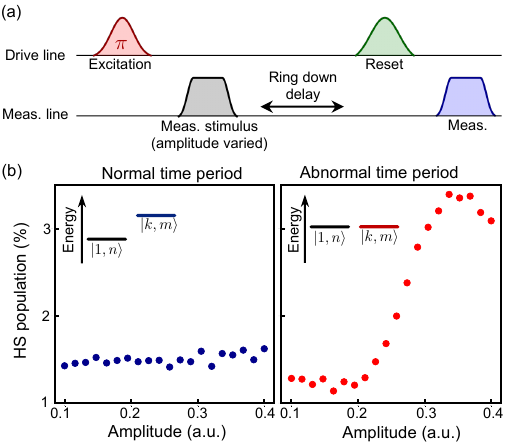}
\caption{
(a) Experimental pulse sequence. The qubit is excited to the $\ket{1}$ state with the $\pi$ pulse, and then a measurement stimulus is transmitted to the readout resonator with a varied amplitude. The qubit $\ket{1}$ state is initialized into the $\ket{0}$ state with the reset operation, and finally, we measure the population of the $\ket{\mathrm{HS}}$ states (states higher than the first excited state). (b) Population of the $\ket{\mathrm{HS}}$ states as a function of the measurement stimulus amplitude. The left (right) panel shows the $\ket{\mathrm{HS}}$ population in the normal (abnormal) time period. We denote the higher dressed states as $\ket{k, m}$ in the insets.
}
\label{f4}
\end{figure}

The mechanism of the MIST is provided in the previous report~\cite{PhysRevLett.117.190503}. We label the energy eigenstates of a qubit-photon system as $\ket{\mathrm{qubit}, \mathrm{resonator}}$. The interaction between the qubit and the resonator imparts a photon-number-dependent energy shift to the bare states. As a result of this energy shift, the dressed state $\ket{1, n}$ can be in resonance with a $\ket{k, m}$ state, where $k$ is an integer larger than 1 and $m$ denotes some photon number. The temporal fluctuations in Fig. \ref{f2} can be explained by the temporal resonant transition between qubit dressed states.

We then varied the number of photons in the readout resonator and measured the HS population. The pulse sequence of the experiment is illustrated in Fig. \ref{f4}(a). The qubit is prepared in the $\ket{1}$ state with the $\pi$ pulse, and we populate the readout resonator with photons using a measurement stimulus. The amplitude of the stimulus is varied, corresponding to injecting different numbers of photons into the resonator. The $\ket{1}$ state is initialized into the $\ket{0}$ state with the reset operation. We finally measure the qubit and estimate the population of the $\ket{\mathrm{HS}}$ state as shown in Fig. \ref{f4}(b) (see Appendix~\ref{est} for the details of the estimation). We interleaved a time series $P_1$ measurement in Fig. \ref{f2} with this experiment, distinguishing between an abnormal (higher error) time period and a normal (lower error) time period.

The result shows that as the amplitude of the measurement stimulus is increased, the $\ket{\mathrm{HS}}$ population also increases during the abnormal time period. Moreover, the population distribution exhibits a resonance-like peak, reaching its maximum value at an amplitude of approximately $0.33$ (in arbitrary units). This peak amplitude is close to the measurement amplitude of $0.28$ (in arbitrary units). Notably, this peak structure is exclusively observed during the abnormal time period, as shown in Fig. \ref{f4}(b). This result suggests that the dressed higher state comes into close proximity with the dressed $\ket{1}$ state only within a specific time interval, as illustrated in the insets.

We can provide only speculations behind the observed temporal resonant transition, but we attribute this to a fluctuating offset charge of the transmon qubit~\cite{PhysRevApplied.20.054008, riste2013millisecond, PhysRevApplied.11.014030, PhysRevApplied.11.024003, PRXQuantum.4.020312}. The higher energy levels of the transmon qubit are sensitive to a change in the offset charge, resulting in a temporal resonant transition between qubit dressed state. The previously reported time scale of the changes in the offset charge~\cite{riste2013millisecond} is consistent with that of the observed instability in this paper, supporting the discussion. We note that though the resonant transition occurs between the first excited state and higher states in this device, the transition between the qubit ground and higher states can be triggered in a system with different qubit parameters or a different measurement amplitude. In addition, in contrast to the prior study~\cite{PhysRevApplied.20.054008}, the switching behavior of the qubit dressed states is observed in a superconducting qubit system where the readout frequency is larger than the qubit transition frequency $\omega_{01}$.

In conclusion, our time-resolved MIST measurement for a superconducting qubit-resonator system suggests anomalous excitation of the qubit beyond the computational subspace during a specific time period. We showed that the MIST shows temporal fluctuation in a variety of time windows ranging from sub-second to a few hundred seconds. The population of the excited state exhibits a peak behavior as a function of the measurement stimulus amplitude only during a specific time interval. These experimental results suggest that resonant transition between the qubit dressed states can be triggered due to a fluctuating offset charge even in a superconducting qubit system coupled with high-frequency resonators.

\begin{acknowledgments}
The authors thank Ted Thorbeck and Matthias Steffen for providing valuable comments on the manuscript. This work was supported by CREST (Nos. JPMJCR20C1, JPMJCR20T2) from JST, Japan; Grant-in-Aid for Scientific Research (S) (No. JP19H05600), Grant-in-Aid for Transformative Research Areas (No. JP22H05114) from JSPS KAKENHI, Japan. 
This work is partly supported by IBM-UTokyo lab.
\end{acknowledgments}

\appendix
\section{TIME SERIES $P_1$ MEASUREMENT}\label{tsp1}
\begin{figure}[t]
\includegraphics{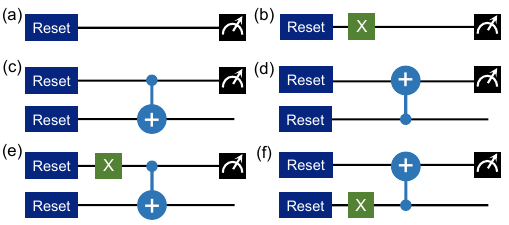}
\caption{
Quantum circuits used in the experiment in Fig.~\ref{f2}. The six circuits were executed in the order (a)$\to$(b)$\to$(c)$\to$(d)$\to$(e)$\to$(f)$\to$(a)$\to\cdots$.
}
\label{f5}
\end{figure}

\begin{figure}[t]
\includegraphics{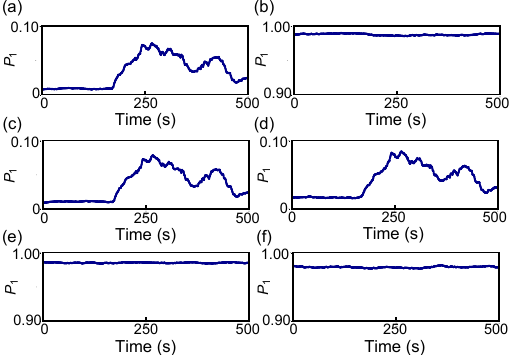}
\caption{
The time series of $P_1$ obtained in Fig.~\ref{f5}. (a) This figure was reproduced from the output in Fig.~\ref{f5}(a), and the same for the other five figures. (c)(d)(e)(f) These were reproduced from the output of the first qubit.
}
\label{f6}
\end{figure}

In the experiment shown in Fig.~\ref{f2}(a), we executed several quantum circuits alternately to capture changes in the error rates in practical quantum computation. The circuits are illustrated in Fig.~\ref{f5}. We incorporated basic operations in quantum computation such as the NOT gate, the CNOT gate, the reset operation, and the measurement in the six circuits, and investigated a temporal increase in the error rates. The six circuits depicted in the figure were executed in the order (a)$\to$(b)$\to$(c)$\to$(d)$\to$(e)$\to$(f)$\to$(a)$\to\cdots$. The time series of $P_1$ from each of the six quantum circuits is shown in Fig.~\ref{f6} with its corresponding letter. When using the CNOT gate, we analyzed the output of the first qubit as depicted in Fig.~\ref{f5}.

In Fig.~\ref{f6}(a), a temporal increase in the error rate is observed from around 200 s to 500 s. In Fig.~\ref{f6}(b), this peak is not evident. This observation aligns with the fact that the in phase component of HS states is positive, and it is converted to 1. As mentioned in the main text, however, the reset operation is unable to initialize the HS states into the ground state, and this excitation leads to a temporal increase in the error rates in other circuits. In Fig.~\ref{f6}(c), we checked the HS excitation of the first qubit results in a higher error in the second qubit due to the CNOT gate. Comparing Fig.~\ref{f6}(a), (c) and (d), we can see that the entanglement with another qubit does not result in the change in the error rates. Notably, though a temporal increase in the error rates is not evident in the first qubit in Fig.~\ref{f6}(e), we observed a temporal increase in the error rates in the second qubit triggered by the HS excitation of the first qubit. In Fig.~\ref{f6}(f), we measure the qubit first excited state, and this results in the error in Fig.~\ref{f6}(a). Figure~\ref{f2}(a) was reproduced from the output in Fig.~\ref{f5}(a). 

\section{ESTIMATION OF HS POPULATION FROM IQ DATA}\label{est}
\begin{figure}[t]
\includegraphics{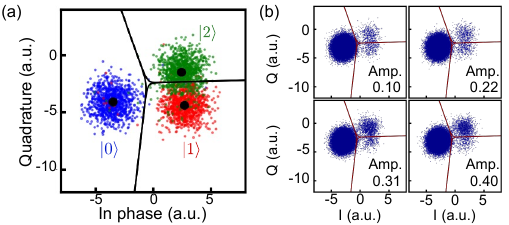}
\caption{
Estimation of the populations of each quantum state from IQ dots. 
(a) Training IQ data and LDA. (b) IQ data obtained in the experiment in Fig.~\ref{f4}(b) and LDA.
}
\label{f7}
\end{figure}

In Fig.~\ref{f7}, we illustrate the procedure for estimating the population of higher energy states (HS). Linear Discriminant Analysis (LDA) was implemented on the IQ data obtained in Fig.~\ref{f1}(b). In Fig.~\ref{f7}(a), the three clusters are classified by LDA, as indicated by the black line. The trained LDA is then used to estimate the population of HS states from the measured IQ data. In Fig.~\ref{f7}(b), IQ dots obtained in the experiment in Fig.~\ref{f4} are presented. Each panel corresponds to the measured signals with a different measurement stimulus amplitude. The amplitude of the measurement stimulus is shown in the inset, and with increasing amplitude, dense signals in the HS region are observed. We applied LDA to convert the IQ dots into the three quantum states and plotted the HS population as a function of the measurement amplitude in Fig.~\ref{f4}(b).

\bibliography{hirasaki_manuscript}

\end{document}